\documentclass[%
superscriptaddress,
reprint,
showpacs,preprintnumbers,
 amsmath,amssymb,
]{revtex4-1}

\usepackage{changes}
\usepackage{upgreek}
\usepackage[T1]{fontenc}
\usepackage[utf8]{inputenc}
\usepackage[english]{babel}

\usepackage[pdftex,unicode,colorlinks,pdfstartview=FitH,linkcolor=black,citecolor=black,%
	pdftitle={Langevin dynamics with spatial correlations as a model for electron-phonon coupling}, %
	pdfauthor={A. Tamm et al} %
	]{hyperref}

\usepackage{graphicx}

\usepackage{amsmath}
\usepackage{braket}
\usepackage{amssymb}

\usepackage{threeparttable}
\usepackage{booktabs}

\usepackage{comment}

\begin{document}

\title{Langevin dynamics with spatial correlations as a model for electron-phonon coupling}

\author{A. Tamm}
\thanks{Corresponding authors: tamm3@llnl.gov, correaa@llnl.gov}
\affiliation{Quantum Simulations Group, Lawrence Livermore National Laboratory, Livermore, California 94550, USA}

\author{M. Caro}
\affiliation{Department of Mechanical Engineering, Virginia Polytechnic Institute, Arlington, VA 22033, USA}

\author{A. Caro}
\affiliation{George Washington University, Ashburn, VA 20147, USA}

\author{G. Samolyuk}
\affiliation{Materials Science and Technology Division, Oak Ridge National Laboratory, Oak Ridge, TN 54321, USA}

\author{M. Klintenberg}
\affiliation{Department of Physics and Astronomy, Uppsala University, 75120 Uppsala, Sweden}

\author{A. A. Correa}
\thanks{Corresponding authors: tamm3@llnl.gov, correaa@llnl.gov}
\affiliation{Quantum Simulations Group, Lawrence Livermore National Laboratory, Livermore, California 94550, USA}

\begin{abstract}
Stochastic Langevin dynamics has been traditionally used as a tool to describe non-equilibrium processes.
When utilized in systems with collective modes, traditional Langevin dynamics relaxes all modes indiscriminately, regardless of their wavelength.
We propose a generalization of Langevin dynamics that can capture a differential coupling between collective modes and the bath, by introducing spatial correlations in the random forces.
This allows modeling the electronic subsystem in a metal as a generalized Langevin bath endowed with a concept of locality, greatly improving the capabilities of the two-temperature model.
The specific form proposed here for the spatial correlations produces physical wavevector- and polarization-dependency of the relaxation produced by the electron-phonon coupling in a solid.
We show that the resulting model can be used for describing the path to equilibration of ions and electrons, and also as a thermostat to sample the equilibrium canonical ensemble.
By extension, the family of models presented here can be applied in general to any dense system, solids, alloys and dense plasmas.
As an example, we apply the model to study the non-equilibrium dynamics of an electron-ion two-temperature Ni crystal.
\end{abstract}

\maketitle


A number of problems in physical sciences involve the motion of particles in 
some sort of medium, for example the Brownian motion of pollen grains in water~\cite{Brown1828}.
If this medium is not reactive, it is usually referred to as a bath~\cite{Tuckerman2010}.
Microscopically, typical baths consist of small particles (e.g water molecules) 
interacting with bigger particles of interest (e.g. proteins). 
The main strategy 
in modeling a bath consists in eliminating irrelevant degrees of freedom and replacing them with a few state variables, such as the temperature of 
the bath.


Under normal conditions, electrons are in equilibrium with the 
ions and respond adiabatically to the motion of ions (e.g. they remain near the 
ground state).
Eliminating the electronic degrees of freedom is very practical because what remains is a problem of classical particles (ions) interacting via an effective 
interatomic potential, i.e. molecular dynamics (MD)~\cite{Alder1957,Alder1959,Car1985}.

This adiabatic elimination is insufficient for many applications of technological 
and scientific importance, such as radiation damage~\cite{DeLaRubia1987,DiazDeLaRubia1991,Zarkadoula2016} 
and laser ablation~\cite{Herrmann1998}. 
For example, when swift ions in a solid induce 
electrons to become excited from their ground state~\cite{Correa2012}; or, 
conversely, when the electrons are heated up by a laser field and transfer their 
energy to the ions~\cite{Ping2009}. 
In these scenarios, systematically ignoring 
the electronic effects becomes increasingly problematic, since fundamental 
physical processes are not accounted for. 

Therefore, practical methods 
aimed at incorporating electron-ion interactions in a realistic way are needed.
One of the most popular and detailed models for capturing ion motion in an 
electronic medium has been the electron-ion two-temperature model \cite{Allen1987} 
coupled with MD (2TM-MD) through Langevin dynamics~\cite{Duffy2007}.

Langevin dynamics, in which each particle is subjected to both a friction and a 
corresponding random force via the fluctuation-dissipation theorem, 
has been historically derived as a model of a bath in 
the context of dilute or liquid systems~\cite{Langevin1908}. 
In the presence of collective modes such as phonons in solids, simple Langevin dynamics as a model of an electronic bath fails in a fundamental way; 
as it damps all motion, including rigid translation of the whole (ions plus electrons) solid.
Moreover, in a recent study~\cite{Tamm2016} we showed that the scalar Langevin model 
introduces the same lifetime for all phonon modes see Fig. 4 in Ref.~\cite{Tamm2016}, regardless of their wave-vector or polarization, which is certainly a severe limitation to study the electron-ion thermalization paths and 
modifications involving the introduction of relative velocities are necessary.

This might not generally be an issue in the case where system properties are 
investigated in equilibrium (e.g. to thermostat or sample a canonical ensemble) or 
if only average ionic aspects are of importance (such as global ionic temperature).
However, measurable microscopic processes occurring in the path to electron-ion 
equilibrium require a good description of the bath, including following certain 
principles, such as global translational invariance and the associated fact that long wavelength modes should be only weakly affected by the coupling with electrons.
\emph{Ad hoc} modifications of Langevin dynamics have been used in the past, but they do not address the fundamental problem of using Langevin dynamics to model non-equilibrium processes in a bath of electrons for a condensed system~\cite{Ma2012}.

Reinforcing the previous argument, experiments have been able to investigate the time-resolved evolution of non-equilibrium phonons excited by hot electrons in pump-probe laser experiments, and observe that not all phonon modes respond equally. 
Diffuse scattering intensity during pump-probe experiments by Chase \emph{et al.}~\cite{Chase2016} demonstrated that the energy transfer from laser-heated electrons to phonon modes near the X and K points in FCC Au proceeds with timescales of \(\sim 2.5~\mathrm{ps}\), faster than for the long wavelength phonon modes.
At the same time, Askerka \emph{et al.}~\cite{Askerka2016} stressed the importance of tensorial and position dependence forms for the electronic friction that is critical in the description of classical ions combined with non-adiabatic electron dynamics of adsorbates at surfaces \cite{HeadGordon1995,Kindt1998,Wodtke2004}.

In the current work we develop a generalization of Langevin dynamics that is able to 
model electron-ion coupling, which is intrinsic to metallic systems. 
The theory is derived on general and simple grounds, such as respecting local translation and rotational invariance and the fluctuation-dissipation theorem; 
but still achieving the required level of complexity needed to model the relaxation of realistic systems. 
The derivation that follows is valid for all systems, 
including crystalline solids, alloys, liquids, and amorphous.


In this model we 
replace the scalar values of friction and random forces over individual particles 
with many-body forces that act in a \emph{correlated} manner over different particles. 
This generalization of Langevin dynamics aims to represent a realistic bath-like interaction with electrons by a friction term~\cite{HeadGordon1995,Kindt1998,Mason2015,Askerka2016} 
(as it is done, for example, for electronic stopping power) 
and a random force (e.g. produced by electronic fluctuations). 
The force on particle \(I\) has three contributions:
\begin{equation}
  \mathbf{f}_I = 
    -\boldsymbol\nabla_I U 
    - \underbrace{\textstyle \sum_J \mathsf{B}_{IJ} \mathbf{v}_J}_{\boldsymbol\upsigma_{I}}
    + \underbrace{\textstyle \sum_J \mathsf{W}_{IJ} \boldsymbol\upxi_J}_{\boldsymbol\upeta_{I}}
  .
  \label{eq:langevin}
\end{equation}

The first term represents the conservative forces, assumed here to be independent 
of the electronic state (implicit dependency of \(U\) on the electronic temperature is ignored here).
The second and third terms are the friction \(\boldsymbol\upsigma\) and random 
forces \(\boldsymbol\upeta\) of the generalized Langevin dynamics, where
random forces are correlated and we make the correlations explicit by the matrix 
and tensor notation:
\begin{equation}
    \textstyle \left< \boldsymbol\upeta_I(t) \boldsymbol\upeta_J(t') \right> = 2 k_\text{B} T_\text{e} \delta(t - t') 
\sum_K\mathsf{W}_{IK}\mathsf{W}^\mathrm{T}_{JK}
    \label{eq:eta2}	
\end{equation}
and where \(\{\boldsymbol\upxi_I\}_I\) is a set of independent white noise 
Gaussian variables with zero mean and no correlation:
\begin{equation}
    \textstyle \left< \boldsymbol\upxi_I(t) \boldsymbol\upxi_J(t') \right> = 2 k_\text{B} T_\text{e} \delta(t - t')\delta_{IJ}
    .
    \label{eq:xi2}
\end{equation}
In addition, \(T_\text{e}\) is the temperature of the bath.
In the special case where matrices are diagonal with scalar elements \(\mathsf{B}_{IJ} = \delta_{IJ}\beta\) 
and \(\mathsf{W}_{IJ} = \delta_{IJ}\sqrt{\beta}\) the standard Langevin 
equation (commonly used in 2TM-MD and as thermostats) is restored. 

The off-diagonal elements in these matrices describe spatial correlations in the system.
This means that the random force \(\boldsymbol\upeta\) acting on a particle in 
a certain direction is not statistically independent of the random force acting on 
other particles and directions. 
In the same way, friction forces \(\boldsymbol\upsigma\) becomes a global property rather than an individual property of each particle.
Friction forces on a particle depend on the velocities of other particles as well.

Based on the fluctuation-dissipation theorem~\cite{Kubo1966}, it can be shown that the friction and random forces need to be related to each other:
\begin{equation}
\textstyle \mathsf{B}_{IJ}=\sum_K\mathsf{W}_{IK}\mathsf{W}^\mathrm{T}_{JK}
\label{eq:BIJ}
\end{equation}

To correct the deficiencies of standard Langevin dynamics 
(i.e. that the collective motions are damped equally),
we propose a modification of the friction term that conserves both linear momentum 
and angular momentum. 
While the principle of conservation of local linear momentum is what controls the qualitative overall shape of inverse-lifetimes of phonon modes with different wavevector (zero at $\Gamma$ and mostly monotonic with $\bf{q}$), the conservation of local angular momentum is the principle that differentiates the lifetime of different polarizations in high symmetry directions.
Instead of following the usual path of defining a friction term, which would require a linear operator decomposition to obtain the random force correlations (Eq.~\ref{eq:eta2}), we start by imposing specific conditions on the random force and derive the matrix \(\mathsf{W}\) from which friction matrix \(\mathsf{B}\) can be obtained through matrix multiplication (Eq.~\ref{eq:BIJ}).

Although random forces are still uncorrelated at different times as in the standard Langevin dynamics, there are now spatial correlations of the forces because the electronic bath has \emph{spatial locality} (i.e. friction forces depend only on relative velocities of nearby ions). 
We require that the random forces do not generate linear or 
angular moments \emph{locally} at any time.
This guiding principle implies a stronger condition than imposing conservation 
of center of mass motion of the whole system, which is the usual \emph{ad hoc} 
method to remove center of mass in equilibrium simulations~\cite{Ogando2002}. 

We propose a recipe to generate the spatially correlated random forces and the 
corresponding friction. 
This is achieved by generating uncorrelated random vectors;
\(\{\boldsymbol\upxi_I\}_I\) which are \emph{initially} assigned to individual 
atoms and follow (component by component) an independent Gaussian distribution,
related to a heat bath at temperature \(T_\text{e}\) (Eq.~\ref{eq:xi2}). 
The \(\boldsymbol\upxi_I\) is then projected on unit vectors connecting neighboring atoms producing a pair decomposition.
Next, each projection is scaled by a weighting factor \(\rho_J(r_{IJ})/\bar{\rho}_I\), which takes into account the influence of atom \(J\) to the site \(I\) (located at distance \(r_{IJ}\)), and summed over neighbors. 
Finally, compensating forces are generated at neighboring atoms by reversing the 
direction of projected and scaled components.
For the radial function \(\rho\) we propose to use atomic electronic densities where the total density at site \(I\) is the 
sum of contributions by its neighbors defined by the radial density ($\bar{\rho}_I = \sum_{J\ne I} \rho_J(r_{IJ})$). 
Therefore, this definition will ensure that the weights at a site will add up to one. 
The use of atomic densities to define a local environment is inspired by the Embedded Atom Method (EAM)~\cite{Daw1984} where they are used as radial functions that define many-body potentials in metallic systems.
Also, the use of electronic density introduces a natural
concept of locality and the relative magnitude of contributions from atoms at different distances. 
Finally, all the pair forces are added up for the total random force on each particle \(\boldsymbol\upeta_I = \sum_J \boldsymbol\upzeta_{IJ}\).

\begin{figure}[t!]
  \includegraphics[width=0.48\textwidth]{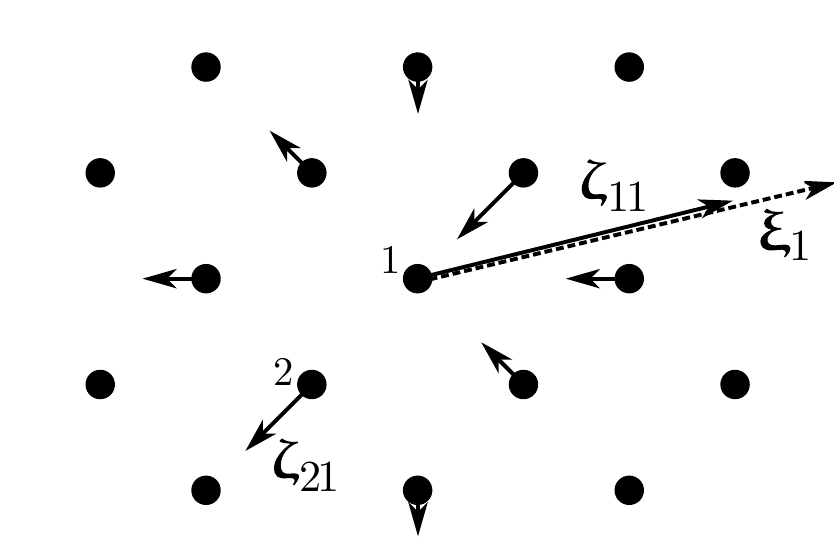}
  \caption{
    Schematic picture describing the partial forces $\boldsymbol\upzeta_{11}$ and $\boldsymbol\upzeta_{21}$ created by the random vector \(\boldsymbol\upxi_1\). 
    Total forces and torques are zero, linear and angular momentum are conserved.
  }
  \label{fig:force}
\end{figure}

In summary, we propose the following form for the pair-projected forces
\begin{equation}
\boldsymbol\upzeta_{IJ} = 
	\begin{cases} 
		\textstyle -\alpha_J
		\frac{\rho_I(r_{IJ})}{\bar{\rho}_J} \mathbf{e}_{IJ} (\boldsymbol\upxi_J \cdot \mathbf{e}_{IJ}), & (I\neq J) \\
		\textstyle \alpha_I \sum_{K\ne I} \frac{\rho_K(r_{IK})}{\bar{\rho}_I} \mathbf{e}_{IK}(\boldsymbol\upxi_I \cdot \mathbf{e}_{IK}), & (I = J)
	\end{cases}
	\label{eq:zetaij}
\end{equation}
where \(\mathbf{e}_{IJ}\) is the unit vector joining atoms \(I\) and \(J\). 
In addition, 
we have introduced the parameter $\alpha_I$ that includes the physics of the coupling 
strength between specific ion types and electrons. 

The projection on \(\mathbf{e}_{IJ}\) is necessary to remove local torque at 
site \(I\) generated by the random vector \(\boldsymbol\upxi_J\). 
In the particular case of an isolated dimer, any component of the random force that is 
perpendicular would become compensated. 

We introduced the concept of locality via a weighting function across neighbors, represented by radial function \(\rho_I\) associated with each ion \(I\). 
Physically, the locality is a property of the electronic system and may be defined by extra knowledge of the electronic system. 
In the current study, the unperturbed atomic electron density (density of isolated atoms) is used for the weighting so that atoms far away will feel the effects of the heat bath at position \(I\) less than atoms closer. 
Also, it acts as a cutoff distance to limit the sums. 
The schematic illustration of the procedure is depicted in Fig. \ref{fig:force}.

With this definition of the random force, the net force and torque of the system 
is zero (exactly, not only in average). 
More importantly, \emph{any} arbitrary partition 
of the system will have this property. 

The elements \(\mathsf{W}_{IJ}\) can be obtained explicitly from the definition of 
\(\boldsymbol\upeta_{I}\) in Eq.~\ref{eq:langevin} and the explicit relation 
with \(\{\boldsymbol\upxi_I\}_I\) in Eq.~\ref{eq:zetaij}.
The tensor elements \(\mathsf{B}_{IJ}\) can be computed by operator multiplication 
(Eq.~\ref{eq:BIJ}).
By construction, this results in a positive-definite symmetric operator \(\mathsf{B}_{IJ} = \mathsf{B}^\mathrm{T}_{JI}\).
Since the operator is definite positive, the second term in Eq.~\ref{eq:langevin} will do negative work.
The friction forces also have the property that the net force and torque of the system is zero. 
(See Supplementary material.)

This algorithm defines a correlation between the components of the random forces 
on individual particles as well as across particles. 
Correspondingly, the set of associated friction forces is linear in the set of 
\emph{all} the velocities. The friction force on a specific particle can 
depend on the velocity of a sufficiently close neighbor.

We also note that there are multiple ways of defining the correlations and 
friction forces still compatible with the requirement of Eq. \ref{eq:eta2}. 
For example, the same conditions could be satisfied by simpler models having a 
friction term and random forces that act only on specific atom pair bonds 
(collection of dimers) as it is done in dissipative particle dynamics~\cite{Hoogerbrugge1992,Espanol1995},
although this would be too constraining compared with the model proposed here 
for solids and condensed systems. In this paper we choose to exploit a pair 
structure in a different way, and we also note that there is an important degree 
of freedom in this model given by the radial functions \(\rho(r)\).

As a test, the model presented above is implemented in \textsc{Lammps} code~\cite{Plimpton1995, LAMMPSFIX} and a Ni crystal is used as an electron-ion system to study three different cases to show its applicability in both non-equilibrium as well as equilibrium and, additionally, as a thermostat.

We use atomic-like density of Ni to construct the weighting function and 
obtain an approximation for the coupling parameters \(\{\alpha_I\}_I\) between the ionic and electronic systems from \emph{ab initio} TDDFT calculations~\cite{Schleife2012}. 
These parameters define the model fully for a specific material and are described
in our previous work~\cite{Tamm2016}.
(Other methods can be utilized to fit or obtain the coupling parameters \emph{ab initio}, 
such as surface hopping~\cite{Tully1990}, adiabatic perturbation theory~\cite{Horsfield2016} or linear response~\cite{Gaspari1972}.)
For our examples, an approximate value of $\alpha^2 = 0.01~\mathrm{eV\, ps/\text{\AA}^2}$ is used and the conservative forces between Ni atoms are defined by EAM~\cite{Daw1984} type potential parametrized by Mishin \emph{et. al}~\cite{Mishin1999} which fairly reproduces the experimental phonon dispersion. 
(Other force fields can be used in this method, empirical or \emph{ab initio}~\cite{Car1985}.)

\begin{figure}
  \begin{center}
    \includegraphics[width=0.48\textwidth]{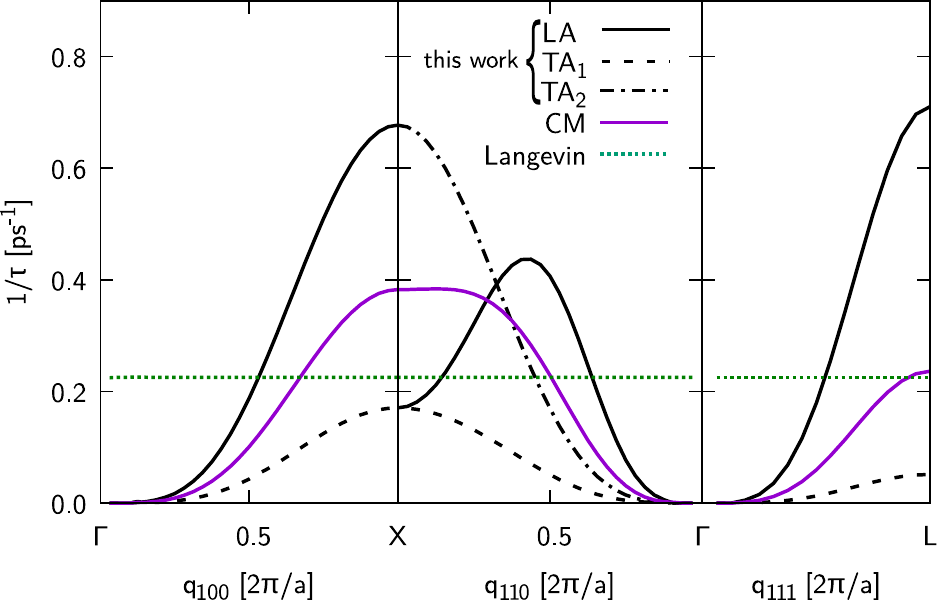}
  \end{center}
  \caption{
    Inverse lifetimes of phonon modes along high symmetry directions in Ni crystal as resulting from our model (black lines) compared to standard Langevin dynamics (green flat line).
    Rotational invariance is related to the factor `\(\mathbf{e}\cdot(\boldsymbol\upxi\cdot\mathbf{e})\)' in Eq.~\ref{eq:zetaij}, and is not enforced if this factor is replaced by a simple non-projected `\(\boldsymbol\upxi\)' (CM-model, see Supplementary material).
    If rotational invariance is not enforced there is still dependence on the wavevector \(q\) but not dependency on the branch (TA or LA) (CM model, purple line). 
    The difference between the lifetimes of TA and LA modes is obtained when the rotational invariance is introduced in the model (full model, black lines). 
  }
  \label{fig:nilife}
\end{figure}

First, we investigate the lifetimes of phonon normal modes in Ni crystal with MD and compare the results with the standard Langevin dynamics. 
The system studied is composed of \(32\times 32\times 32\) FCC 
conventional cells (131072 atoms) with periodic boundary conditions. 
In this simulation the normal modes are excited individually with a small amplitude 
to reduce the effect of anharmonicity on the phonon lifetimes. 
The electronic bath is \emph{held} at constant 
\(T_\text{e} = 0\) for the purpose of these lifetime calculations. The amplitude 
of each phonon mode with wavevector \(\mathbf{q}\) is monitored during the 
simulation and fitted with an exponent to obtain the lifetime. 
The result for Ni is shown on Fig.~\ref{fig:nilife}. 
The main trend that we observe as a result of our model is that the  lifetimes near the \(\Gamma\) point in reciprocal space 
(long wavelength) are longer than near the Brillouin zone border 
(short wavelength), exhibiting a certain characteristic shape. 
At the same time longitudinal modes (LA) have shorter lifetimes than transverse modes (TA) near \(\Gamma\). 
Tests show that the resulting shape is influenced by the choice of the weighting function \(\rho\). 
This can be exploited in subsequent work to improve over our first ansatz of using atomic densities as radial functions. 
The quality of the radial function can be assessed by comparison to phonon lifetimes obtained from DFT linear perturbation theory calculations~\cite{Gaspari1972} or $\mathbf{q}$-resolved non-equilibrium pump-probe experiments~\cite{Chase2016}.

A key goal of the model is proven here, namely, that obtained lifetimes have a 
wavevector \emph{and} polarization dependence. In particular, electrons 
do not damp phonons of long wavelength. Furthermore, we observe that 
different branches have different lifetimes. 
This rich behavior not only emerges 
from the principles utilized to construct the forces in the Langevin model, but it is also in qualitative agreement with more elaborate quantum mechanical 
treatments of the coupling~\cite{Tamm2016}.
The perfect crystal helps in the interpretation of friction forces (second term in Eq. 1).
Although they are not necessarily antiparallel to the velocities, under small oscillations these forces and velocities are antiparallel in the normal coordinates.
The corresponding friction coefficients are zero for translations and rotations of the crystal as a whole.


\begin{figure}
  \includegraphics[width=0.48\textwidth]{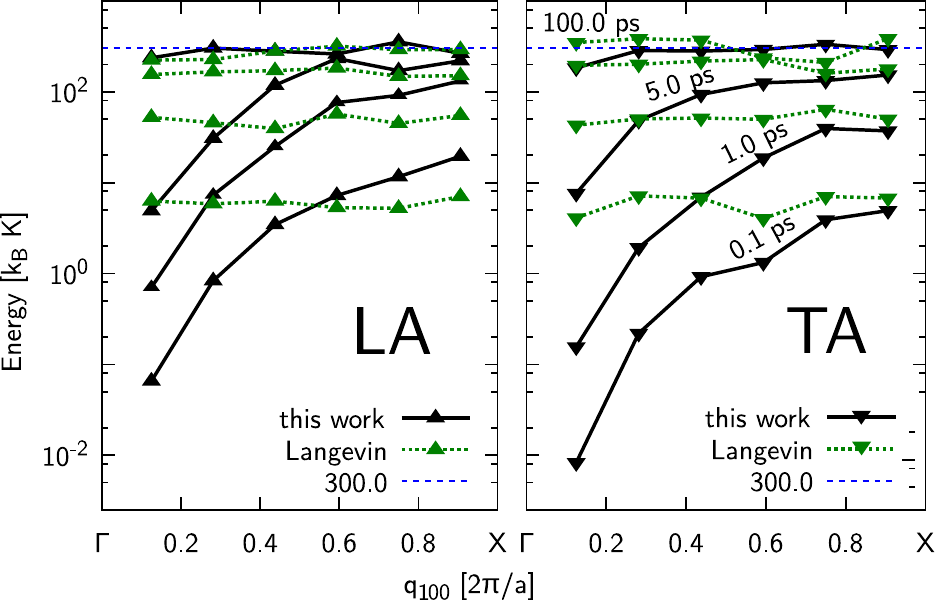}
  \caption{
	Energy per mode at times \(0.1\), \(1\), \(5\), and \(100~\mathrm{ps}\) for the modified Langevin (this work, black lines) and for standard Langevin (dotted green). 
	Since in the latter the energy transferred to phonons is independent of the phonon mode, there is an incidental near fulfillment of the equipartition of energy between modes at intermediate times. 
	In the model proposed in this work, equipartition between modes 
	(an equal energy per mode) is only a property of the final equilibrium state.
	We give energy in units of \(k_\text{B} \mathrm{K}\) for reference, 
	since a system in equilibrium will have an energy \(k_\text{B} T\) per mode.
	In all cases the final state corresponds to equilibrium at \(300~\mathrm{K}\), the target set by \(T_\text{e}\).
	  }
  \label{fig:energymode}
\end{figure}

Next, we equilibrated the ionic system starting at $0~\mathrm{K}$, 
with electronic bath held at $T_\text{e} = 300~\mathrm{K}$. 
The amplitudes of individual modes were extracted from atomic coordinates and 
velocities. We have plotted the energy of phonon modes along \(\langle 100\rangle\) 
direction in Fig.~\ref{fig:energymode}. 
It can be seen that in our model the modes are excited at different rates, depending
on the strength of the coupling, whereas the traditional 2TM-MD model (standard Langevin dynamics) excites all the modes at the same rate, which shows that it will give a poor microscopic representation 
of the electron-phonon equilibration process. 
Also, with the 2TM-MD all modes 
are incidentally in approximate equipartition relative to each other at all times within the simulation, 
whereas with our model the energy is \emph{transiently} distributed reflecting the \(\mathbf{q}\)-dependence of the electron-phonon coupling, which also respects the LA and 
TA differences. Moreover, the 2TM-MD is often used to study the evolution of 
non-equilibrium processes, and having the incorrect equilibration process questions 
the applicability of such studies to the non-equilibrium path to stable or 
metastable final states. 


Lastly, we analyzed the velocity distribution of ions after the system has equilibrated. Our results agree with the expected Maxwell-Boltzmann distribution, therefore showing that the model acts as a thermostat, similarly with the standard Langevin formalism.


In conclusion, we have derived a novel model to describe the interaction of ions 
in a solid with electrons modeled as a heat bath, by expanding Langevin dynamics 
through spatial correlations of the random forces and a corresponding friction 
term consistent with the fluctuation-dissipation theorem~\cite{Kubo1966}. 
We also give a procedure to define the parameters in the model from higher order 
methods such as DFT and non-adiabatic TDDFT.

The model was applied to study the phonon lifetimes in FCC Ni. 
It was seen that the lifetimes depend both on the \(\mathbf{q}\)-vector 
and polarization of the phonon mode, which is not the case in the original Langevin formulation.
Next, we looked at the non-equilibrium evolution of the phonon population following hot electrons at \(300~\mathrm{K}\).
We observed that phonons with wavevectors close to the Brillouin zone boundaries are
excited first due to the stronger coupling, and that energy equipartition between the modes is achieved asymptotically in agreement with recent experimental observations~\cite{Chase2016}.

Finally, our model is computationally simple, since friction and random forces 
are consistent without the cost of matrix decomposition methods and can be 
readily used to study various processes that are far from equilibrium, such as 
two-temperature conditions generated by fast laser heating. 
We propose this model for the general study of systems where ions and electrons interact non-adiabatically while maintaining a classical atomic framework. 
Although this theory was applied to electrons in a metallic system, 
a similar construction is likely necessary for any medium or 
bath whose collective motion is \emph{attached} to the Brownian particles that 
move inside it, where a bath with the concept of locality naturally arises.


This work was performed under the auspices of the U.S. Department of Energy by 
Lawrence Livermore National Laboratory under Contract DE-AC52-07NA27344. 
This work was supported as part of the Energy Dissipation to Defect Evolution (EDDE), an Energy Frontier Research Center funded by the US Department of Energy, Office of Science, Basic Energy Sciences.

\bibliography{article}

\begin{thebibliography}{10}
\expandafter\ifx\csname url\endcsname\relax
  \def\url#1{\texttt{#1}}\fi
\expandafter\ifx\csname urlprefix\endcsname\relax\def\urlprefix{URL }\fi
\providecommand{\bibinfo}[2]{#2}
\providecommand{\eprint}[2][]{\url{#2}}

\bibitem{Brown1828}
\bibinfo{author}{Brown, R.}
\newblock \bibinfo{title}{Xxvii. a brief account of microscopical observations
  made in the months of june, july and august 1827, on the particles contained
  in the pollen of plants; and on the general existence of active molecules in
  organic and inorganic bodies}.
\newblock \emph{\bibinfo{journal}{The Philosophical Magazine}}
  \textbf{\bibinfo{volume}{4}}, \bibinfo{pages}{161--173}
  (\bibinfo{year}{1828}).

\bibitem{Tuckerman2010}
\bibinfo{author}{Tuckerman, M.~E.}
\newblock \emph{\bibinfo{title}{Statistical Mechanics: Theory and Molecular
  Simulation (Oxford Graduate Texts)}} (\bibinfo{publisher}{Oxford University
  Press}, \bibinfo{year}{2010}).

\bibitem{Alder1957}
\bibinfo{author}{Alder, B.~J.} \& \bibinfo{author}{Wainwright, T.~E.}
\newblock \bibinfo{title}{{Phase transition for a hard sphere system}}.
\newblock \emph{\bibinfo{journal}{J. Chem. Phys.}}
  \textbf{\bibinfo{volume}{27}}, \bibinfo{pages}{1208--1209}
  (\bibinfo{year}{1957}).

\bibitem{Alder1959}
\bibinfo{author}{Alder, B.~J.} \& \bibinfo{author}{Wainwright, T.~E.}
\newblock \bibinfo{title}{{Studies in molecular dynamics. General method}}.
\newblock \emph{\bibinfo{journal}{J. Chem. Phys.}}
  \textbf{\bibinfo{volume}{31}}, \bibinfo{pages}{459} (\bibinfo{year}{1959}).

\bibitem{Car1985}
\bibinfo{author}{Car, R.} \& \bibinfo{author}{Parrinello, M.}
\newblock \bibinfo{title}{Unified approach for molecular dynamics and
  density-functional theory}.
\newblock \emph{\bibinfo{journal}{Phys. Rev. Lett.}}
  \textbf{\bibinfo{volume}{55}}, \bibinfo{pages}{2471--2474}
  (\bibinfo{year}{1985}).

\bibitem{DeLaRubia1987}
\bibinfo{author}{{De La Rubia}, T.~D.}, \bibinfo{author}{Averback, R.~S.},
  \bibinfo{author}{Benedek, R.} \& \bibinfo{author}{King, W.~E.}
\newblock \bibinfo{title}{{Role of thermal spikes in energetic displacement
  cascades}}.
\newblock \emph{\bibinfo{journal}{Phys. Rev. Lett.}}
  \textbf{\bibinfo{volume}{59}}, \bibinfo{pages}{1930--1933}
  (\bibinfo{year}{1987}).

\bibitem{DiazDeLaRubia1991}
\bibinfo{author}{{Diaz De La Rubia}, T.} \& \bibinfo{author}{Guinan, M.~W.}
\newblock \bibinfo{title}{{New mechanism of defect production in metals: A
  molecular-dynamics study of interstitial-dislocation-loop formation in
  high-energy displacement cascades}}.
\newblock \emph{\bibinfo{journal}{Phys. Rev. Lett.}}
  \textbf{\bibinfo{volume}{66}}, \bibinfo{pages}{2766--2769}
  (\bibinfo{year}{1991}).

\bibitem{Zarkadoula2016}
\bibinfo{author}{Zarkadoula, E.}, \bibinfo{author}{Samolyuk, G.},
  \bibinfo{author}{Xue, H.}, \bibinfo{author}{Bei, H.} \&
  \bibinfo{author}{Weber, W.~J.}
\newblock \bibinfo{title}{{Effects of two-temperature model on cascade
  evolution in Ni and NiFe}}.
\newblock \emph{\bibinfo{journal}{Scr. Mater.}} \textbf{\bibinfo{volume}{124}},
  \bibinfo{pages}{6--10} (\bibinfo{year}{2016}).

\bibitem{Herrmann1998}
\bibinfo{author}{Herrmann, R. F.~W.}, \bibinfo{author}{Gerlach, J.} \&
  \bibinfo{author}{Campbell, E. E.~B.}
\newblock \bibinfo{title}{{Ultrashort pulse laser ablation of silicon: an MD
  simulation study}}.
\newblock \emph{\bibinfo{journal}{Appl. Phys. A Mater. Sci. Process.}}
  \textbf{\bibinfo{volume}{66}}, \bibinfo{pages}{35--42}
  (\bibinfo{year}{1998}).

\bibitem{Correa2012}
\bibinfo{author}{Correa, A.~A.}, \bibinfo{author}{Kohanoff, J.},
  \bibinfo{author}{Artacho, E.}, \bibinfo{author}{S\'anchez-Portal, D.} \&
  \bibinfo{author}{Caro, A.}
\newblock \bibinfo{title}{Nonadiabatic forces in ion-solid interactions: The
  initial stages of radiation damage}.
\newblock \emph{\bibinfo{journal}{Phys. Rev. Lett.}}
  \textbf{\bibinfo{volume}{108}}, \bibinfo{pages}{213201}
  (\bibinfo{year}{2012}).

\bibitem{Ping2009}
\bibinfo{author}{Ping, Y.} \emph{et~al.}
\newblock \bibinfo{title}{Warm dense matter created by isochoric laser
  heating}.
\newblock \emph{\bibinfo{journal}{High Energy Density Physics}}
  \textbf{\bibinfo{volume}{6}}, \bibinfo{pages}{246 -- 257}
  (\bibinfo{year}{2010}).
\newblock \bibinfo{note}{ICHED 2009 - 2nd International Conference on High
  Energy Density Physics}.

\bibitem{Allen1987}
\bibinfo{author}{Allen, P.~B.}
\newblock \bibinfo{title}{Theory of thermal relaxation of electrons in metals}.
\newblock \emph{\bibinfo{journal}{Phys. Rev. Lett.}}
  \textbf{\bibinfo{volume}{59}}, \bibinfo{pages}{1460--1463}
  (\bibinfo{year}{1987}).

\bibitem{Duffy2007}
\bibinfo{author}{Duffy, D.~M.} \& \bibinfo{author}{Rutherford, A.~M.}
\newblock \bibinfo{title}{{Including the effects of electronic stopping and
  electron–ion interactions in radiation damage simulations}}.
\newblock \emph{\bibinfo{journal}{J. Phys. Condens. Matter}}
  \textbf{\bibinfo{volume}{19}}, \bibinfo{pages}{016207}
  (\bibinfo{year}{2007}).

\bibitem{Langevin1908}
\bibinfo{author}{Langevin, P.}
\newblock \bibinfo{title}{{Sur la th\'{e}orie du mouvement brownien}}.
\newblock \emph{\bibinfo{journal}{C. R. Acad. Sci. Paris}}
  \textbf{\bibinfo{volume}{146}}, \bibinfo{pages}{530--533}
  (\bibinfo{year}{1908}).

\bibitem{Tamm2016}
\bibinfo{author}{Tamm, A.} \emph{et~al.}
\newblock \bibinfo{title}{{Electron-phonon interaction within classical
  molecular dynamics}}.
\newblock \emph{\bibinfo{journal}{Phys. Rev. B}} \textbf{\bibinfo{volume}{94}},
  \bibinfo{pages}{024305} (\bibinfo{year}{2016}).

\bibitem{Ma2012}
\bibinfo{author}{Ma, P.-W.}, \bibinfo{author}{Dudarev, S.~L.} \&
  \bibinfo{author}{Woo, C.~H.}
\newblock \bibinfo{title}{Spin-lattice-electron dynamics simulations of
  magnetic materials}.
\newblock \emph{\bibinfo{journal}{Phys. Rev. B}} \textbf{\bibinfo{volume}{85}},
  \bibinfo{pages}{184301} (\bibinfo{year}{2012}).

\bibitem{Chase2016}
\bibinfo{author}{Chase, T.} \emph{et~al.}
\newblock \bibinfo{title}{{Ultrafast electron diffraction from non-equilibrium
  phonons in femtosecond laser heated Au films}}.
\newblock \emph{\bibinfo{journal}{Appl. Phys. Lett.}}
  \textbf{\bibinfo{volume}{108}} (\bibinfo{year}{2016}).

\bibitem{Askerka2016}
\bibinfo{author}{Askerka, M.}, \bibinfo{author}{Maurer, R.~J.},
  \bibinfo{author}{Batista, V.~S.} \& \bibinfo{author}{Tully, J.~C.}
\newblock \bibinfo{title}{{Role of Tensorial Electronic Friction in Energy
  Transfer at Metal Surfaces}}.
\newblock \emph{\bibinfo{journal}{Phys. Rev. Lett.}}
  \textbf{\bibinfo{volume}{116}}, \bibinfo{pages}{1--5} (\bibinfo{year}{2016}).

\bibitem{HeadGordon1995}
\bibinfo{author}{{Head Gordon}, M.} \& \bibinfo{author}{Tully, J.~C.}
\newblock \bibinfo{title}{{Molecular dynamics with electronic frictions}}.
\newblock \emph{\bibinfo{journal}{J. Chem. Phys.}}
  \textbf{\bibinfo{volume}{103}}, \bibinfo{pages}{10137--10145}
  (\bibinfo{year}{1995}).

\bibitem{Kindt1998}
\bibinfo{author}{Kindt, J.~T.}, \bibinfo{author}{Tully, J.~C.},
  \bibinfo{author}{Head-Gordon, M.} \& \bibinfo{author}{Gomez, M.~A.}
\newblock \bibinfo{title}{Electron-hole pair contributions to scattering,
  sticking, and surface diffusion: Co on cu(100)}.
\newblock \emph{\bibinfo{journal}{The Journal of Chemical Physics}}
  \textbf{\bibinfo{volume}{109}}, \bibinfo{pages}{3629--3636}
  (\bibinfo{year}{1998}).

\bibitem{Wodtke2004}
\bibinfo{author}{Wodtke, A.~M.}, \bibinfo{author}{Tully, J.~C.} \&
  \bibinfo{author}{Auerbach, D.~J.}
\newblock \bibinfo{title}{Electronically non-adiabatic interactions of
  molecules at metal surfaces: Can we trust the {B}orn–{O}ppenheimer
  approximation for surface chemistry?}
\newblock \emph{\bibinfo{journal}{International Reviews in Physical Chemistry}}
  \textbf{\bibinfo{volume}{23}}, \bibinfo{pages}{513--539}
  (\bibinfo{year}{2004}).

\bibitem{Mason2015}
\bibinfo{author}{Mason, D.}
\newblock \bibinfo{title}{{Incorporating non-adiabatic effects in embedded atom
  potentials for radiation damage cascade simulations}}.
\newblock \emph{\bibinfo{journal}{J. Phys. Condens. Matter}}
  \textbf{\bibinfo{volume}{27}}, \bibinfo{pages}{145401}
  (\bibinfo{year}{2015}).

\bibitem{Kubo1966}
\bibinfo{author}{Kubo, R.}
\newblock \bibinfo{title}{The fluctuation-dissipation theorem}.
\newblock \emph{\bibinfo{journal}{Reports on Progress in Physics}}
  \textbf{\bibinfo{volume}{29}}, \bibinfo{pages}{255} (\bibinfo{year}{1966}).

\bibitem{Ogando2002}
\bibinfo{author}{Ogando, E.}, \bibinfo{author}{Caro, M.} \&
  \bibinfo{author}{Caro, A.}
\newblock \bibinfo{title}{Reference systems for computational free energy
  calculations of binary solutions: role of the constrained center of mass
  motion}.
\newblock \emph{\bibinfo{journal}{Computational Materials Science}}
  \textbf{\bibinfo{volume}{25}}, \bibinfo{pages}{297 -- 304}
  (\bibinfo{year}{2002}).

\bibitem{Daw1984}
\bibinfo{author}{Daw, M.~S.} \& \bibinfo{author}{Baskes, M.~I.}
\newblock \bibinfo{title}{{Embedded-atom method: Derivation and application to
  impurities, surfaces, and other defects in metals}}.
\newblock \emph{\bibinfo{journal}{Phys. Rev. B}} \textbf{\bibinfo{volume}{29}},
  \bibinfo{pages}{6443--6453} (\bibinfo{year}{1984}).

\bibitem{Hoogerbrugge1992}
\bibinfo{author}{Hoogerbrugge, P.~J.} \& \bibinfo{author}{Koelman, J. M. V.~A.}
\newblock \bibinfo{title}{{Simulating Microscopic Hydrodynamic Phenomena with
  Dissipative Particle Dynamics}}.
\newblock \emph{\bibinfo{journal}{Europhys. Lett.}}
  \textbf{\bibinfo{volume}{19}}, \bibinfo{pages}{155--160}
  (\bibinfo{year}{1992}).

\bibitem{Espanol1995}
\bibinfo{author}{Espanol, P.} \& \bibinfo{author}{Warren, P.}
\newblock \bibinfo{title}{{Statistical Mechanics of Dissipative Particle
  Dynamics.}}
\newblock \emph{\bibinfo{journal}{EPL (Europhysics Lett.}}
  \textbf{\bibinfo{volume}{30}}, \bibinfo{pages}{191--196}
  (\bibinfo{year}{1995}).

\bibitem{Plimpton1995}
\bibinfo{author}{Plimpton, S.}
\newblock \bibinfo{title}{{Fast Parallel Algorithms for Short-Range Molecular
  Dynamics}}.
\newblock \emph{\bibinfo{journal}{J. Comput. Phys.}}
  \textbf{\bibinfo{volume}{117}}, \bibinfo{pages}{1--19}
  (\bibinfo{year}{1995}).

\bibitem{LAMMPSFIX}
\bibinfo{title}{{LAMMPS}: {L}arge-scale {A}tomic/{M}olecular massively
  {P}arallel {S}imulator is a classical molecular dynamics code
  \url{http://lammps.sandia.gov}, utilized here in combination with {USER-EPH}
  a custom developed package for parameterized electron-phonon coupling
  \url{http://github.com/LLNL/USER-EPH}}.

\bibitem{Schleife2012}
\bibinfo{author}{Schleife, A.}, \bibinfo{author}{Draeger, E.~W.},
  \bibinfo{author}{Kanai, Y.} \& \bibinfo{author}{Correa, A.~A.}
\newblock \bibinfo{title}{{Plane-wave pseudopotential implementation of
  explicit integrators for time-dependent Kohn-Sham equations in large-scale
  simulations}}.
\newblock \emph{\bibinfo{journal}{J. Chem. Phys.}}
  \textbf{\bibinfo{volume}{137}}, \bibinfo{pages}{0--9} (\bibinfo{year}{2012}).

\bibitem{Tully1990}
\bibinfo{author}{Tully, J.~C.}
\newblock \bibinfo{title}{Molecular dynamics with electronic transitions}.
\newblock \emph{\bibinfo{journal}{The Journal of Chemical Physics}}
  \textbf{\bibinfo{volume}{93}}, \bibinfo{pages}{1061--1071}
  (\bibinfo{year}{1990}).

\bibitem{Horsfield2016}
\bibinfo{author}{Horsfield, A.~P.}, \bibinfo{author}{Lim, A.},
  \bibinfo{author}{Foulkes, W. M.~C.} \& \bibinfo{author}{Correa, A.~A.}
\newblock \bibinfo{title}{Adiabatic perturbation theory of electronic stopping
  in insulators}.
\newblock \emph{\bibinfo{journal}{Phys. Rev. B}} \textbf{\bibinfo{volume}{93}},
  \bibinfo{pages}{245106} (\bibinfo{year}{2016}).

\bibitem{Gaspari1972}
\bibinfo{author}{Gaspari, G.~D.} \& \bibinfo{author}{Gyorffy, B.~L.}
\newblock \bibinfo{title}{Electron-phonon interactions, $d$ resonances, and
  superconductivity in transition metals}.
\newblock \emph{\bibinfo{journal}{Phys. Rev. Lett.}}
  \textbf{\bibinfo{volume}{28}}, \bibinfo{pages}{801--805}
  (\bibinfo{year}{1972}).

\bibitem{Mishin1999}
\bibinfo{author}{Mishin, Y.}, \bibinfo{author}{Farkas, D.},
  \bibinfo{author}{Mehl, M.~J.} \& \bibinfo{author}{Papaconstantopoulos, D.~A.}
\newblock \bibinfo{title}{{Interatomic potentials for monoatomic metals from
  experimental data and ab initio calculations}}.
\newblock \emph{\bibinfo{journal}{Phys. Rev. B}} \textbf{\bibinfo{volume}{59}},
  \bibinfo{pages}{3393--3407} (\bibinfo{year}{1999}).

\end{thebibliography}

\onecolumngrid
\appendix 
\section{Supplementary material}

Equation of motion of the proposed Langevin dynamics with spatial correlations is
\begin{equation}
  \mathbf{f}_I = 
    -\boldsymbol\nabla_I U 
    - \underbrace{\sum_J \mathsf{B}_{IJ} \mathbf{v}_J}_{\boldsymbol\upsigma_{I}}
    + \underbrace{\sum_J \mathsf{W}_{IJ} \boldsymbol\upxi_J}_{\boldsymbol\upeta_{I}}
  ,
\end{equation}
where \(\boldsymbol{\upsigma}_I\) and \(\boldsymbol\upeta_I\) are the friction and random forces respectively. 
Subindices denote particles.

We start by defining the matrix that produces the spatial correlations of the random forces \(\boldsymbol\upeta_I\) when starting from uncorrelated \(\boldsymbol\upxi_J\), \(\boldsymbol\upeta_I = \sum_J \boldsymbol\upzeta_{IJ}\), and \(\boldsymbol\upzeta_{IJ} = \mathsf{W}_{IJ} \boldsymbol\upxi_J\) (no summation):
 \begin{equation}
  \mathsf{W}_{IJ} =
  \begin{cases}
      \displaystyle -\alpha_J \frac{\rho_I(r_{IJ})}{\bar{\rho}_J} \mathbf{e}_{IJ}\mathbf{e}_{IJ} & (I \neq J)\\
      & \\
    \displaystyle \alpha_I \sum_{K \ne I} \frac{\rho_K(r_{IK})}{\bar{\rho}_I} \mathbf{e}_{IK}\mathbf{e}_{IK} & (I = J) 
  \end{cases},
\end{equation}
where \(\rho_I\) are radially decaying functions for atom \(I\) (such as atomic electron density of atom \(I\))
and \(\mathbf{e}_{IJ}\) and \(r_{IJ}\) are the unit vector and distance between two ions at positions \(\mathbf{r}_I\) and \(\mathbf{r}_J\). 
(\(\mathbf{e}_{IJ}\)\(r_{IJ} = \mathbf{r}_I - \mathbf{r}_J\).)
Dyadic notation is used. 

Applying the fluctuation-dissipation theorem we obtain the friction (drag) forces by multiplication \(\mathsf{B}_{IJ}=\sum_K\mathsf{W}_{IK}\mathsf{W}^\mathrm{T}_{JK}\). 
Therefore, the correlations in the friction is given by:
\begin{equation}
  \displaystyle
  \mathsf{B}_{IJ} = 
  \begin{cases}
  \begin{split}
    -\sum_{K \ne I} \alpha_I^2 \frac{\rho_J(r_{IJ})\rho_K(r_{IK})}{\bar{\rho}_I^2} 
    (\mathbf{e}_{IJ} \cdot \mathbf{e}_{IK})\mathbf{e}_{IJ}\mathbf{e}_{IK} 
    -\sum_{K \ne J} \alpha_J^2 \frac{\rho_I(r_{IJ})\rho_K(r_{JK})}{\bar{\rho}_J^2}
    (\mathbf{e}_{IJ} \cdot \mathbf{e}_{JK})\mathbf{e}_{IJ}\mathbf{e}_{JK} + \\
    +\sum_{K \ne I \land K \ne J} \alpha_K^2 \frac{\rho_I(r_{IK})\rho_J(r_{JK})}{\bar{\rho}_K^2}
    (\mathbf{e}_{IK} \cdot \mathbf{e}_{JK})\mathbf{e}_{IK}\mathbf{e}_{JK}
  \end{split} & (I \neq J) \\
   & \\
    \begin{split}
    \sum_{K \ne I} \sum_{L \ne I} \alpha_I^2 \frac{\rho_K(r_{IK}) \rho_L(r_{IL})}{\bar{\rho}_I^2} 
    (\mathbf{e}_{IK} \cdot \mathbf{e}_{IL}) \mathbf{e}_{IK}\mathbf{e}_{IL} + 
    \sum_{K \ne I} \alpha_K^2 \frac{\rho_I^2(r_{IK})}{\bar{\rho}_K^2} \mathbf{e}_{IK}\mathbf{e}_{IK} 
  \end{split} & (I = J)
  \end{cases}.
\end{equation}
After applying the operator \(\mathsf{B}\) on velocities we get the friction forces \(\boldsymbol\upsigma_I = \sum_J \mathsf{B}_{IJ} \mathbf{v}_J\):
\begin{subequations}
\begin{alignat}{4}
\boldsymbol\upsigma_I =  
    \sum_{K \ne I} & \left(\alpha_I^2 \tfrac{\rho_K^2(r_{IK})}{\bar{\rho}_I^2}  +  
    \alpha_K^2 \tfrac{\rho_I^2(r_{IK})}{\bar{\rho}_K^2}\right) 
     \mathbf{e}_{IK}((\mathbf{v}_I - \mathbf{v}_K)\cdot\mathbf{e}_{IK}) + \\ 
	& + \sum_{K \ne I} \sum_{L \ne I \land L \ne K} \alpha_I^2 \frac{\rho_K(r_{IK})\rho_L(r_{IL})}{\bar{\rho}_I^2}
    (\mathbf{e}_{IK} \cdot \mathbf{e}_{IL})\mathbf{e}_{IK}( (\mathbf{v}_I - \mathbf{v}_K)\cdot\mathbf{e}_{IL}) + \\
    & \quad\quad + \sum_{K \ne I} \sum_{L \ne I \land L \ne K} \alpha_L^2 \frac{\rho_I(r_{IL})\rho_K(r_{KL})}{\bar{\rho}_L^2}
    (\mathbf{e}_{IL} \cdot \mathbf{e}_{KL})\mathbf{e}_{IL}( (\mathbf{v}_K - \mathbf{v}_L)\cdot\mathbf{e}_{KL})
\end{alignat}
\end{subequations}
This form shows that the drag force is only a function of velocity differences, but it is not very efficient
due to the double summations.
In the implementation, the calculation could be split into two parts, with an auxiliary set \(\mathbf{w}_I\): 
\begin{equation}
  \mathbf{w}_I = 
		\sum_J \mathsf{W}^\mathrm{T}_{JI} \mathbf{v}_J =
    \sum_{J \neq I} \alpha_I \frac{\rho_J(r_{IJ})}{\bar{\rho}_I} \mathbf{e}_{IJ} (\mathbf{v}_I - \mathbf{v}_J) \cdot \mathbf{e}_{IJ}
\end{equation}
\begin{equation}
  \boldsymbol\upsigma_I = \sum_J \mathsf{W}_{IJ} \mathbf{w}_J
	= \sum_{J \neq I} \mathbf{e}_{IJ}
		\left(
			   \alpha_I \frac{\rho_J(r_{IJ})}{\bar{\rho}_I} \mathbf{w}_I
			 - \alpha_J \frac{\rho_I(r_{IJ})}{\bar{\rho}_J} \mathbf{w}_J
		\right)\cdot \mathbf{e}_{IJ}
\end{equation}

Any arbitrary partition \(\mathcal{P}\) of the system will have these properties: 
\begin{equation}
\sum_{I\in\mathcal{P}} \sum_{J\in\mathcal{P}} \boldsymbol\upzeta_{IJ} = \mathbf{0}
\end{equation}
\begin{equation}
\sum_{I\in\mathcal{P}} \mathbf{r}_I \times \sum_{J\in\mathcal{P}}\boldsymbol\upzeta_{IJ} = \mathbf{0}, 
\label{eq:localangular}
\end{equation}

Globally,
\begin{equation}
\sum_{I} \boldsymbol\upeta_{I} = \mathbf{0}, \quad   \sum_{I} \boldsymbol\upsigma_{I} = \mathbf{0}
\end{equation}
\begin{equation}
\sum_{I} \mathbf{r}_I \times \boldsymbol\upeta_{I} = \mathbf{0}, \quad \sum_{I} \mathbf{r}_I \times \boldsymbol\upsigma_{I} = \mathbf{0}
\label{eq:globalangular}
\end{equation}
(at all times, not only in average)

In the specific case, where the angular parts are omitted (substituting $\mathbf{e}_{IJ}\mathbf{e}_{IJ}$ with identity matrix) as in the `CM-model' in Fig.~2. 
The random forces are generated by:
 \begin{equation}
  \mathsf{W}_{IJ} =
  \begin{cases}
      \displaystyle -\alpha_J \frac{\rho_I(r_{IJ})}{\bar{\rho}_J} & (I \neq J)\\
      & \\
    \displaystyle \alpha_I \sum_{K \ne I} \frac{\rho_K(r_{IK})}{\bar{\rho}_I} & (I = J)
  \end{cases}.
\end{equation}
and the friction force will be 
\begin{subequations}
\begin{align}
\boldsymbol\upsigma_I = 
    \sum_{K \ne I} \alpha_I^2 \frac{\rho_K(r_{IK})}{\bar{\rho}_I}  
    (\mathbf{v}_I - \mathbf{v}_K) \, + & \\
     + \sum_{K \ne I} \alpha_K^2 \frac{\rho_I^2(r_{IK})}{\bar{\rho}_K^2} 
    (\mathbf{v}_I - \mathbf{v}_K) & + 
    \sum_{K \ne I} \sum_{L \ne I \land L \ne K} \alpha_L^2 \frac{\rho_I(r_{IL})\rho_K(r_{KL})}{\bar{\rho}_L^2}
    (\mathbf{v}_K - \mathbf{v}_L).
\end{align}
\end{subequations}
This simplified model will damp the rotations of a whole system, and Eq.~\ref{eq:localangular} and Eq.~\ref{eq:globalangular} will not hold.

\end{document}